\documentclass[aip,amsmath,amssymb,preprint]{revtex4-1}

\usepackage{graphicx}

\begin{document}


\title{Thermal emittance measurements of a cesium potassium antimonide photocathode}

\author{Ivan Bazarov}
\author{Luca Cultrera}
\author{Adam Bartnik}
\author{Bruce Dunham}
\author{Siddharth Karkare}
\author{Yulin Li}
\author{Xianghong Liu}
\author{Jared Maxson}
\author{William Roussel}
\noaffiliation
\affiliation{Cornell Laboratory for Accelerator-based Sciences and Education, Cornell University, Ithaca, NY 14853, USA}

\begin{abstract}
Thermal emittance measurements of a $\mathrm{CsK_2Sb}$ photocathode at several laser wavelengths are presented. The emittance is obtained with a solenoid scan technique using a high voltage dc photoemission gun. The thermal emittance is $0.56\pm0.03$\,mm\,mrad/mm(rms) at 532\,nm wavelength. The results are compared with a simple photoemission model and found to be in a good agreement.
\end{abstract}

\maketitle

Photocathodes play an increasingly important role in present and future accelerators requiring increasingly smaller emittance and higher current beams utilizing photoemission for electron generation. Not only high quantum efficiency (QE), but also small thermal (intrinsic) emittance and sub-picosecond photoemission response are generally required. $\mathrm{CsK_2Sb}$ is well-known from streak camera applications and the photomultiplier tube industry to have prompt response and high QE in a convenient (visible) spectral range. However, thermal emittance, the other critical parameter for accelerators, has never been reported in the literature. We present such measurements obtained using a solenoid scan technique with a high voltage dc photoemission gun for $\mathrm{CsK_2Sb}$ excited with 532, 473, and 405\,nm light. The results are compared with a simple theory and found in excellent agreement when using the values for band gap energy and electron affinity reported in the literature.

\begin{figure}[bhtp]
\includegraphics[width=\columnwidth]{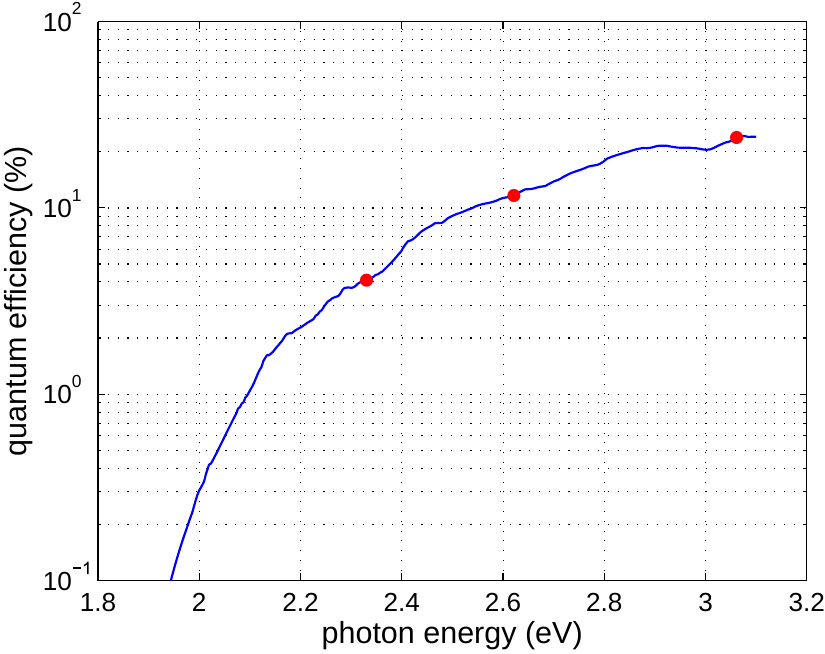}
\caption{QE vs.~photon energy.\label{fig:qe}}
\end{figure}
$\mathrm{CsK_2Sb}$ photocathodes are grown on heavily doped Si substrates in a dedicated chamber and subsequently translated into the high voltage dc gun\cite{bruce} equipped with a load-lock system. Details of the photocathode preparation chamber and cathode recipe are available elsewhere\cite{luca}. Photocathodes with QE of 5-7\,\% in the green are routinely produced. Fig.~\ref{fig:qe} shows QE vs.~the photon energy for the photocathode used in these measurements. Specific photon energies used in this experiment are indicated by the circles.

\begin{figure}[htbp]
\includegraphics[width=\columnwidth]{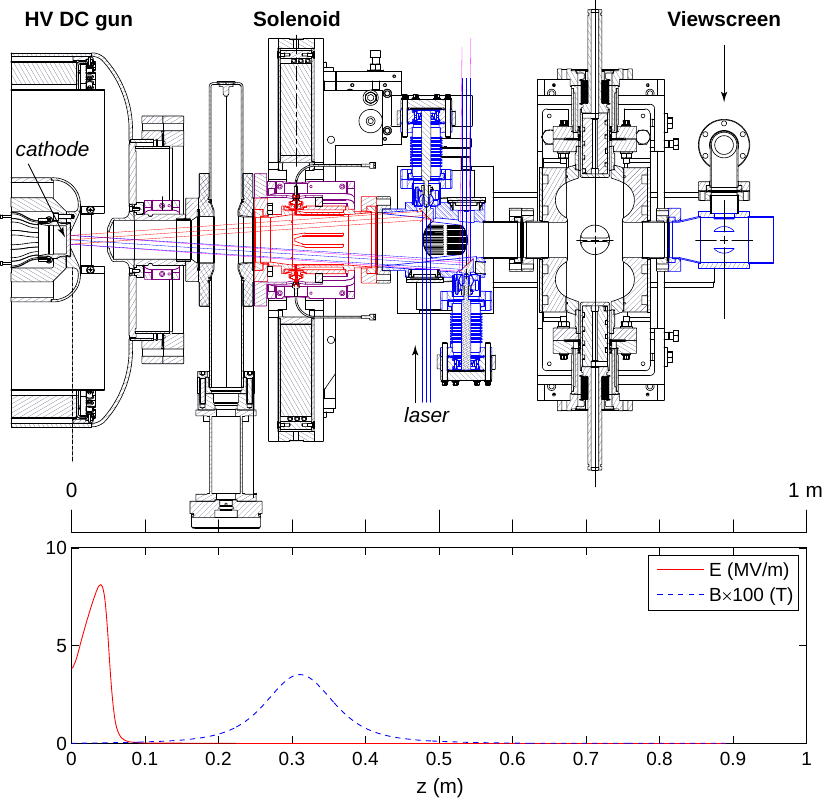}
\caption{Experimental setup and fields.\label{fig:a1}}
\end{figure}
The experimental beamline used to measure thermal emittance is shown in Fig.~\ref{fig:a1}. The relevant beamline elements are the dc gun, the solenoid, and the Chemical Vapor Deposition (CVD) diamond viewscreen positioned at 0.89\,m from the photocathode. The solenoid current is varied and the beam size at the screen is recorded. The data is later fitted to obtain the emittance. The axial electric and magnetic fields are shown in Fig.~\ref{fig:a1} for 350\,kV gun voltage and 4.2\,A solenoid current. We employ dc lasers to deliver several $\mu$A of beam current. The effects of the space charge are negligible as a result.

For the solenoid scan technique to work, it is critical to know the transfer matrices $\mathbf{R} = \mathbf{R}(i \rightarrow f)$, which relate the particle's transverse coordinate $x$ and divergence $dx/dz$ from before to after the region of interest:
\begin{equation}
\left( \begin{array}{c}
x \\
\frac{dx}{dz} \end{array} \right)_{\!\!f} = \,\mathbf{R}
\left( \begin{array}{c}
x \\
\frac{dx}{dz} \end{array} \right)_{\!\!i}.
\end{equation}
The use of a single solenoid implies coupling of $x, y$ transverse planes, which requires more complicated treatment for beams with asymmetry and generally $4 \times 4$ size transfer matrices. Additionally, electric and magnetic fields overlap in a small region of our beamline and need to be properly accounted for. We found an efficient way to address this by obtaining an exact analytical solution for the $2 \times 2$ transfer matrix in the Larmor frame for the overlapping constant $E,B$ fields (both along longitudinal or $z$ direction) of length $\Delta z$:
\begin{equation}
\mathbf{R}(\Delta z) = \left( \begin{array}{cc}
C & S\frac{\beta_i \gamma_i}{b}\\
-S\frac{b}{\beta_f \gamma_f} & C\frac{\beta_i \gamma_i}{\beta_f \gamma_f} \end{array} \right).
\end{equation}
Here, $\gamma$ and $\beta$ are Lorentz factors before (subscript $i$) and after (subscript $f$) the field region. $C \equiv \cos(\Delta\theta)$ and $S \equiv \sin(\Delta\theta)$ for Larmor angle $\Delta\theta \equiv \theta_f - \theta_i = (b/a) \ln(\gamma_f(1+\beta_f)/\gamma_i(1+\beta_i))$. Normalized field strengths are $a \equiv eE/mc^2$ and $b \equiv eB/2mc$. Lorentz factors before and after the field region are related by $\gamma_f = \gamma_i + a \Delta z$. The fundamental constants here are the electron mass $m$, charge $e$, and the speed of light $c$.
The edge effect matrix (for rising and falling edges of $E$) is given by
\begin{equation}
\mathbf{R}_{\mathrm{edge},i,f} = \left( \begin{array}{cc}
1 & 0\\
\mp\frac{a}{2 \beta_{i,f}^2 \gamma_{i,f}} & 1 \end{array} \right).
\end{equation}
The upper sign is taken with $i$ subscript for the entrance (rising) edge $\mathbf{R}_{\mathrm{edge},i}$, and the other choice for the exit (falling) edge. The full matrix from the cathode to the screen location (and the overall Larmor angle) is obtained by matrix multiplication of small $\Delta z$ slices of $E(z)$ and $B(z)$ fields using $\mathbf{R}_{\mathrm{edge},f} \mathbf{R}(\Delta z) \mathbf{R}_{\mathrm{edge},i}$ for each individual interval (except right at the cathode, which does not include a rising $E$ edge). Upon applying the counter-rotation to the screen image by the Larmor angle $\theta$, the motion in $x$ and $y$ planes becomes decoupled and the solenoid scan analysis proceeds in a usual manner, e.g.~see \cite{ivan}. The matrix formalism and the overall Larmor angle were verified both experimentally and through the particle tracking and were found to be in excellent agreement.

\begin{figure}[bhtp]
\includegraphics[width=\columnwidth]{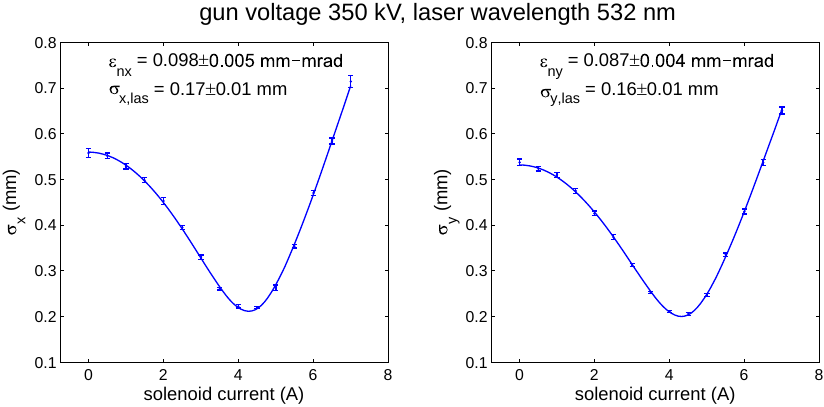}
\caption{Solenoid scan and fit example.\label{fig:fit}}
\end{figure}
Upon rotation by the Larmor angle, each viewscreen image was processed\cite{scubeex} to extract rms beam size in both planes. The rms beam size vs.~the solenoid current is used to determine the beam emittance, see Fig.~\ref{fig:fit} for an example of the solenoid scan fit. The emittance was measured at 3 difference gun voltages: 110\,kV, 230\,kV, and 350\,kV and was found to be the same within the uncertainty of the measurement. The 350\,kV gun voltage corresponds to about 4\,MV/m at the photocathode, which results in a Schottky correction to the work function of less than 0.1\,eV.

The laser beam was 1:1 imaged onto the cathode after passing through a circular aperture. To account for possible photocathode QE variations and slight differences in laser delivery path to the gun, we use the results of the solenoid scan fit which returns the initial electron beam size at the location of the cathode in addition to the beam emittance. These values are in a good agreement with those obtained by measuring the laser image at a CCD camera positioned at the same distance from the imaging lens as the gun photocathode.

\begin{figure}[htbp]
\includegraphics[width=\columnwidth]{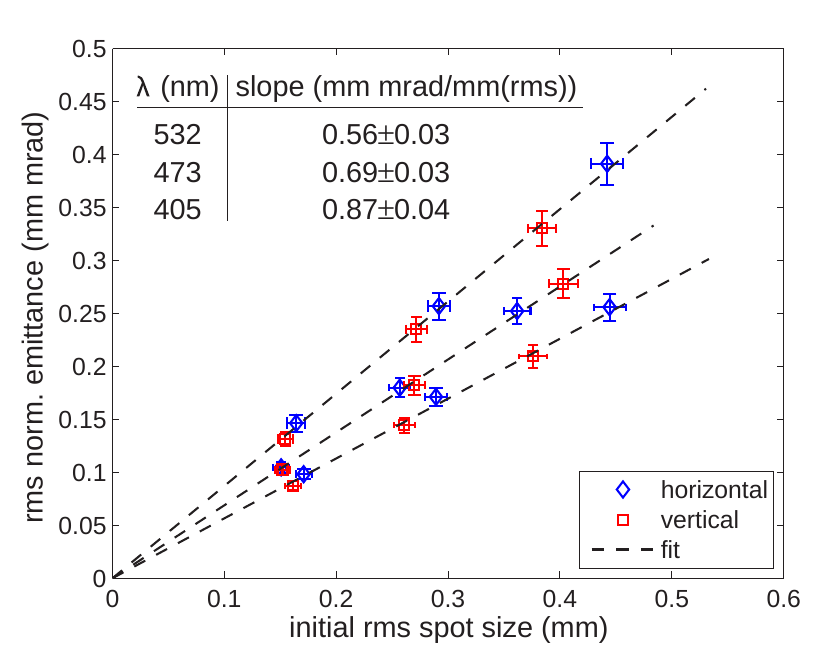}
\caption{Thermal emittance for different wavelengths.\label{fig:summary}}
\end{figure}
The summary of the results is shown in Fig.~\ref{fig:summary} for the three wavelengths (the gun voltage is 230\,kV). The slope of normalized rms emittance with the rms laser spot size is indicated for each of the laser wavelength.

\begin{figure}[hbtp]
\includegraphics[width=\columnwidth]{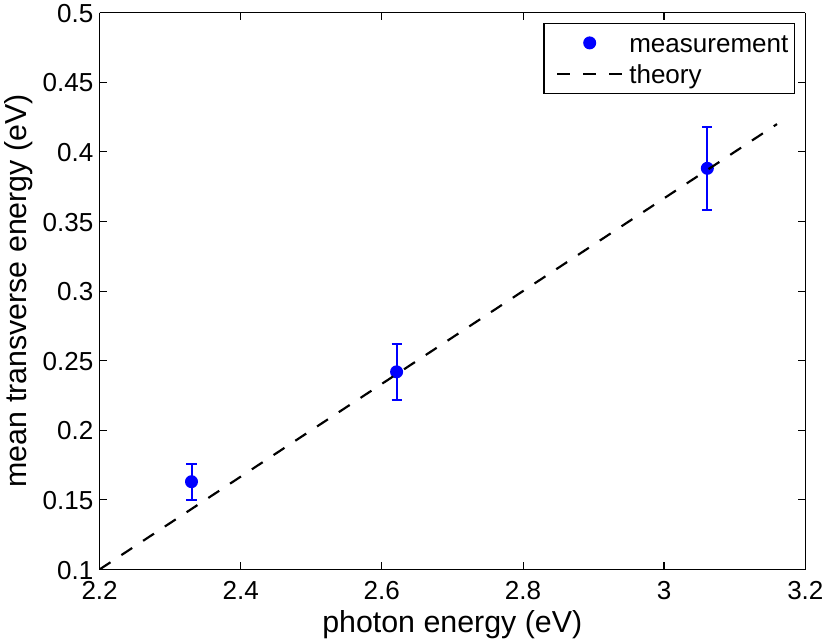}
\caption{Measured MTE and Eq.~\ref{eq:mte} vs.~photon energy.\label{fig:mte}}
\end{figure}
Mean Transverse Energy (MTE) is often used as a figure of merit to characterize photocathode thermal emittance independent of the laser spot size. The rms normalized emittance at the cathode is given by
\begin{equation}
\epsilon_{nx} = \sigma_x \frac{\sqrt{\left<p_x^2\right>}}{mc} = \sigma_x \sqrt{\frac{\mathrm{MTE}}{mc^2}},
\end{equation}
with $\left<p_x^2\right>$ the transverse momentum variance, and $\mathrm{MTE} = \left<E_x\right> + \left<E_y\right>$ the average transverse kinetic energy of photoemitted electrons, a quantity readily measurable with electron energy analyzers. Here, isotropic photoemission is assumed, i.e.~$\mathrm{MTE} = 2\left<E_x\right> = \left<p_x^2\right>/m$. Additionally, for electrons emitted with an excess energy $\left<E\right> = 3\left<E_x\right>$ into the vacuum hemisphere, one writes $\mathrm{MTE} = 2\left<E\right>/3$. For a given energy gap $E_g$ and positive electron affinity $E_a$,  the simplest model assumes that the photoelectrons are distributed uniformly with kinetic energies between $h\nu - (E_g + E_a)$ and 0, or $\left<E\right> = (h\nu - (E_g + E_a))/2$ and therefore\cite{dowell}
\begin{equation}
\mathrm{MTE} = \frac{h\nu-(E_g + E_a)}{3}.\label{eq:mte}
\end{equation}
Fig.~\ref{fig:mte} shows MTE obtained from our data and compared with Eq.~\ref{eq:mte} for energy gap $E_g = 1.2$\,eV and electron affinity $E_a = 0.7$\,eV\cite{ghosh}.

In summary, the measurements of transverse emittance for $\mathrm{CsK_2Sb}$ are presented. The results are found in good agreement with a simple photoemission model. We conclude that this material is very suitable for applications requiring both high QE, low thermal electron emittance, and a prompt response.

This work is supported by NSF DMR-0807731. Two of the authors (S.K. and partially L.C.) are supported by DOE DE-SC0003965.

\end{document}